\documentclass[12pt]{iopart}

\usepackage{harvard}
\usepackage{graphicx}

\newcommand{\bc}{\begin{center}}
\newcommand{\ec}{\end{center}}
\begin{document}
\title{Visualizing Astrophysical N-body Systems}


\author{John Dubinski}
\address{Department of Astronomy and Astrophysics, 50 St. George Street, University of Toronto, Toronto,
ON, Canada M5S 3H4}
\ead{dubinski@astro.utoronto.ca; url:http://www.galaxydynamics.org}



\begin{abstract}
I begin with a brief history of N-body simulation and visualization
and then go on to describe various methods for creating images and animations of
modern simulations in cosmology and galactic dynamics.
These techniques are incorporated into a specialized particle visualization
software library called MYRIAD that is designed to render images within large
parallel N-body simulations as they run.
I present several case studies that explore the application of these
methods to animations of star clusters, 
interacting galaxies and cosmological structure formation.
\end{abstract}




\maketitle

\section{Introduction}

One of the great challenges in numerical astrophysics right now
is the simulation of the formation and evolution of galaxies from the initial
conditions prescribed by the $\Lambda$-CDM cosmological model
\cite{wmap08}.
This model specifies the cosmological parameters and initial power spectrum
of density fluctuations and presents an accurate set of initial conditions
for the universe emerging from the Big Bang.
The difficult task at hand is
numerically solving the physical equations governing gravity, gas
dynamics, and radiation to understand the details of how
the initial state transforms 
into a universe containing the large-scale structure 
and galactic population revealed by observations.
Modern simulations employ both Lagrangian (particle or N-body) and
Eulerian (mesh) methods either separately or as hybrids to follow the
evolution of the dark matter and gas and its transformation into galaxies
and their stellar populations.

Advances in supercomputing power combined with code development and
new efficient algorithms during the past two decades have created marvelous
datasets with increasing resolution and realism for comparison to 
the real universe.  Current simulations now
contain billions of particles or comparable numbers of mesh
elements.  These datasets are usually studied like real observations 
by analyzing the various gross properties and statistics of galaxy-like objects 
that form in the simulations including masses, scale-lengths, shapes etc. 
Direct visualization of various physical fields such as the 
density, pressure and
temperature are also a
powerful way to convey the dynamical evolution.  When image time sequences
are joined together to make animations, 
we develop new intuition into the evolution
of structures as well as see the relative importance of competing physical
processes in driving structure formation.  
Visualizations are also important tools for conveying the
complexities of nonlinear dynamical equations to an non-expert audience
including other scientists and the general public at large.

In this paper, I discuss some techniques for the visualization 
of gravitational N-body
simulations as they arise in both cosmological models and experiments in
galactic dynamics.  Historically, advances in N-body methods have gone hand
in hand with visualization and indeed the motivation for doing the first N-body
simulations of galaxies were to reproduce the apparent dynamical features
in the beautiful photographic images of spiral 
galaxies and interacting systems.  In \S
2, I begin by reviewing the historical development of methods of
N-body visualization.
In \S 3, I discuss various methods for visualizing different types of
particle system with some discussion of the application of basic 3D graphics.
In \S 4, I present some case studies to
illustrate the use of these methods taken from the GRAVITAS project,
an exploration of the science and art of 
N-body systems, and then conclude with some remarks on future goals of
achieving astronomical photorealism.

\section{A Brief History of N-body Simulation and Visualization}

A complete history of the development of computing and visualization of
gravity cannot be presented here but it is worth looking at some critical
events and their importance for advancement and understanding in
astrophysics.
The fascination with the motion of heavenly bodies stretches back to
antiquity and modern physics begins here.  
There is an interwoven story 
of both computation and visualization that is told in trying to
understand these motions.

Archimedes is said to have constructed a device that could
trace the movement of the sun, moon based on the
heliocentric models of the day and could be used to
make predictions of the timing of eclipses \cite{arc02}.
Such a mechanical device was
actually retrieved from an ancient Roman wreck near Antikythera dated from
the 2nd century BCE and
while the purpose of the mechanism is still debated it was likely an
analogue computing device that could be used to visualize the motion of the
sun, moon and planets and predict their future positions \cite{fre06}.
Mechanical armillery spheres and astrolabes 
were later developed independently by Chinese and Arab astronomers
to computing and visualize celestial motions.

The advance of the heliocentric Copernican world view lead to
Kepler's discoveries of the planetary laws of motion.   His reduction 
of Tycho's detailed observations of Mars path across the sky showed the 
motion to be the result 
the combined effect of the Earth and Mars on their
own elliptical orbits.   The motions of the planets were therefore
visualized as confocal ellipses - geometrical curves that could 
be traced out on paper.

In the {\em Principia Mathematica}, Newton clearly defined and solved 
the two-body problem with the new laws of mechanics and gravity. 
Clockwork mechanisms called orreries (after the Earl of Orrery who 
first commissioned
one to be built) displayed the motions of the planets in the solar system
as predicted by Newton's classical mechanics.    These early visualizations
translated Newton's impenetrable expert work of mathematics of the time
into something that a wider audience could easily understand.  
However, Newton himself realized
the computational difficulty in solving the more general N-body
problem in the context of the solar system and stated:

\begin{quotation}
The orbit of any planet depends on the combined motion of all 
the planets, not to mention the action of all these on each other.  
But to consider simultaneously all these causes of motion and to define 
these motions by exact laws allowing of convenient calculation exceeds, 
unless I am mistaken, the force of the entire human intellect. \cite{new34}
(p. 574)
\end{quotation}

He could not foresee the development of electronic computing that would 
eventually permit the approximate (but accurate) solution 
of the general equations of the N-body problem
through numerical methods in the 20th century.\footnote{Ironically, modern
computers function under the principles of quantum mechanics and so
understanding of the full complexity of Newton's N-body problem
required the discovery of new physics to construct the machines
which perhaps did require the force
of the entire human intellect.}  

With the development of large reflecting telescopes the universe was
opened up to the discovery of the nebulae some of which were the galaxies.
Lord Rosse's sketches of the spiral patterns in the Whirlpool Galaxy (M51), 
were highly suggestive of a swirling, rotating object.
Lord Rosse noted: 
\begin{quotation}
The sketches \ldots convey a pretty accurate idea of
the peculiarities of structure which have gradually become known to us: in
many of the nebulae they are very remarkable, and seem even to indicate the
presence of dynamical laws we may perhaps fancy to be almost within our
grasp. -The Earl of Rosse \citeyear{ros50}  
\end{quotation}
The connection of spiral structure to Newtonian
gravity would come later.
Chamberlin \citeyear{cha1901} first conjectured that the spiral nebulae
might the remnants of the gravitational tidal interactions of colliding
bodies. At that time the concept of a galaxy was not defined so he
guessed the colliding objects might be stars.
Lindblad
recognized the connection of these processes to stellar dynamics 
and struggled with the problem in the middle of the 20th century.
One pioneering advance was carried out by Erik Holmberg
\citeyear{hol41} in an ingenious table top experiment 
to study interacting galaxies.

In the era before general purpose electronic computing, Holmberg
\citeyear{hol41} conceived an
experiment which used light-bulbs as proxies for gravitating point masses.
Since the intensity of light falls as an inverse square law then the
cumulative intensity measured at a given position could be used as a proxy
for the net gravitational force from the system.  He laid out two disk
configurations consisting of 37 particles to represent two interacting
spiral galaxies.  Forces were ``computed" by measuring the intensity from the
$x$ and $y$ directions at the positions of each light bulb and recorded
meticulously in
a notebook.  After a force sweep, the positions of the lightbulbs were
changed according to their assigned velocity and measured acceleration over
a timestep.  After each reconfiguration, the process was repeated
and so a numerical integration of the motion of the particles could proceed
relatively quickly using light intensity as an analogue computational
device.  In this way, he provided the first early evidence that tidal
encounters of disk galaxies can induce spiral structure using simulation
and visualization.  

%

The history of modern computation is interwoven with 
the solution of the N-body problem and it's visualization.
The first computers like the ENIAC were general purpose
and programmable and so applicable to many problems.  Computation of
trajectories for ballistic firing tables strangely occupied as much as 25\%
of their time \cite{ree52}!  
The application of computing the N-body problem to the fields of
stellar dynamics emerged in the early 1960's with applications to globular
star clusters \cite{vHo60} and galaxy clusters \cite{aar63}.
Many of the issues related to numerical accuracy and computational
efficiency that are still relevant today
were discussed by Aarseth \& Hoyle \citeyear{aar64}.
Direct N-body methods using typically several hundred particles
through the 60's and 70's were used to illustrate various problems in the
evolution of globular star clusters,
collisionless gravitational collapse \cite{pee70}, disk stability
\cite{ost73}, cosmological
structure formation \cite{aar79} and interacting galaxies \cite{whi78}.
Particle-mesh methods derived from techniques employed in plasma physics
\cite{mil68,mil70}
permitted much larger numbers of particles and were used at this time to
explore the stability of galactic disks \cite{hoh71} and visualizations
clearly demonstrated the growth of intrinsic spiral and bar 
instabilities (See \cite{sel87} for
a discussion of these methods).
The seminal work by the Toomre brothers \citeyear{too72} 
on interacting galaxies demonstrated the power of well-crafted figures
to express a physical idea.  Using restricted 3-body calculations, they
clearly demonstrated the origin of
tails and bridges seen in close pairs of galaxies as the result of
gravitational tidal interactions rather than some other process.  This work
was also animated using super 8 film and is perhaps the earliest animated
example of a physically based simulation of galaxies \cite{too06}. Eneev \etal
\citeyear{ene73} also  did similar work and created an film animation in
the Soviet Union \cite{sun06}.

The growth of computing power through the 80's combined with algorithmic
developments transformed the way the N-body problem was solved and
visualization played a continuing major role in the 
understanding the processes in galactic dynamics and
cosmology.  
The emerging cold
dark matter cosmology  revealed a rich pattern of clustering that
could be compared directly to maps of the large-scale structure
\cite{dav85}.
In galactic dynamics, the application of particle-mesh methods
\cite{sel87} and
the Barnes-Hut (BH) tree algorithm \citeyear{bar86} increased particle number
from hundreds to tens of thousands permitting accurate representation of
self-consistent disk galaxies.
Barnes' stunning simulation of a galaxy group with 64K particles
\cite{bar89}
was transformed into one of the first detailed animations of
self-consistent galaxy dynamics and had great impact both in scientific
understanding as well as public fascination with astronomy.
The combination of smoothed-particle hydrodynamics \cite{luc77} with the tree
algorithm by Hernquist \citeyear{her89} further increased the realism
associated with the simulation of galaxies.

As the modern methods of computer graphics, rendering and animation
explosively emerged in the 90's they saw applications in astrophysics.
The production of the IMAX film Cosmic Voyage in 1996 \cite{cos96} 
featured animations
of cosmological structure formation from the work of Summers \etal
\citeyear{sum96} and SPH galaxy collisions from the work of Mihos \&
Hernquist \citeyear{mih94} and were created with the help of PIXAR studios and
visual artists such as Cox at the NCSA \cite{cox96}.  This work raised the
bar for the standard of astrophysical animation.

At the present time, the parallelization of N-body algorithms along with
specialized hardware has increased the numbers of particles in simulations
into the billions.  There are now many parallelized N-body codes based on
different algorithms including treecodes
\cite{sal91,war95,dub96a,spr01,wad04}, PM methods \cite{mac98}
along with new PM-tree hybrids
\cite{dub04,spr05}.   These codes running on large parallel
supercomputers now permit simulations with as many as
10 billion particles \cite{spr05a}.  Also notable, is the development of the
GRAvity PipE (GRAPE) specialized hardware \cite{mak03} 
that permits integration of the
{\em direct} N-body problem for up to a million particles.
Visualization of these enormous
datasets has become increasingly challenging from a data management and
computational point of view.

Concurrently, many particle visualization packages have been developed with
different features depending on the type of simulation and task.  
Some notable specialized open source 
interactive packages are Starsplatter \cite{wel00}, 
Tipsy \cite{qui97}, Partiview \cite{lev03} and SPLASH \cite{pri07}.
Many N-body visualizers also convert their data into 2D and 3D density fields
and use a data analysis and visualization package like IDL to make 
colored maps or volumetric renderings or simply develop their own 
specialized code using open source graphics libraries such as OpenGL.


In the next section, I will describe some general techniques that are used
to render large particle datasets including methods for
creating density fields with different viewing geometries along with
methods of adding color and optimizing brightness and contrast to enhance
detail.

\section{Methods}

\subsection{Rendering particle distributions}

The first N-body simulations were represented graphically as dots on X-Y
plots.  
While this was sufficient when particle numbers 
were less than a few thousand, as particle numbers increased figures of galaxy
encounters turned into black featureless blobs.  Images of real galaxies
are representations of the combined luminosity of hundreds of billions of
stars integrated along the line of sight.  
(We ignore the obscuring effects of interstellar dust for now.)
The quantity observed is the surface brightness.
A similar interpretation can be applied to dark matter simulations
where the simple mass density is represented as an equivalent
luminosity density.  In some cases it is also useful to represent 
the particles as individual points of light such as the rendering of a star 
cluster and we discuss how this can be done realistically as well.

The procedure for estimating the surface density from a particle
distribution is straightforward.  For orthographic (parallel) 
X-Y projections, one simply
specifies a 2D rectangular domain subdivided in pixels and bins the particle
mass in their given pixel according to their position coordinates to create
a density field.  A particle system is a random
subsample of some smooth density field so this procedure is equivalent to
integrating the 3D density down a given line of sight to produce the
surface density.  Surface density maps of this kind can then be converted
into a suitable image format e.g., FITS in astronomy and manipulated within
an astronomical image viewing software such as SAOImage DS9  or IDL to 
view grayscale or pseudo-colour renditions.

In low density regions where particle sampling is sparse it is sometimes
desirable to use interpolation schemes to smooth the density field either
in 2D or 3D.  Convolution with a smoothing function with a single
scale length (e.g., a Gaussian) 
can be carried out quickly using the convolution theorem and
Fast Fourier transforms.   The use of real astronomical point spread functions
(PSF) that include diffraction spikes introduces interesting
visual effects as we discuss below.

The interpolation scheme of smoothed particle hydrodynamics (SPH) provides a
convenient method for estimating the density field from a discrete particle
distribution (see Price \citeyear{pri07} for a thorough
discussion in application to SPH simulations of star formation).  
The density within any pixel (or voxel) 
is determined as as sum over particle densities weighted by 
a kernel function.  An advantage of SPH interpolation is that
smoothing length is adaptive so particles in lower density regions have
larger smoothing lengths.  Variable smoothing lengths therefore permit a
better estimate of the density in sparsely sampled regions than convolution
with a PSF function with a single length scale.   
SPH smoothed images provide the best images for quantitative analysis.

When particle number densities are relatively high i.e., hundreds of
particles per pixel, both of these methods provide similar estimates of the
density and appear similar in visualizations.  In low particle density regions, 
the SPH method will be more accurate but individual particles will be 
smeared out into large-sized blobs many pixels across.
Convolution with a single PSF reveals a transition from a 
smooth density field to a system of discrete particles.
In the creation of animations, both methods can be used to represent
evolving galaxies and cosmological dark matter structures.  The choice of
smoothing is usually a matter of taste.  SPH interpolated systems can appear
overly smoothed and hide dynamical details that may be revealed 
by moving particles while the discrete look of systems convolved with 
a single function may appear too noisy to others.  
For systems such as star clusters, convolution
with an astronomical PSF containing diffraction spikes and the inclusion of
stellar colors can create images that resemble real astronomical images.

\subsection{Basic 3D Computer Graphics}

The calculation of orthographic surface density images assumes 
that rays are parallel to the line of sight.
Sometimes it is interesting to place the camera nearby or immersed within a
particle system thus introducing visual perspective.   Systems containing 
structure on many scales such as seen in simulations of cosmological 
structure formation can then be appreciated more fully.  A camera can
also be dynamic moving along a trajectory with a changing viewing
direction to create animations.  
Animations of rotations and  fly-throughs
give a sense of the relative size and shape of structures in 3
dimensions.

\begin{figure}
\includegraphics[width=15.5cm]{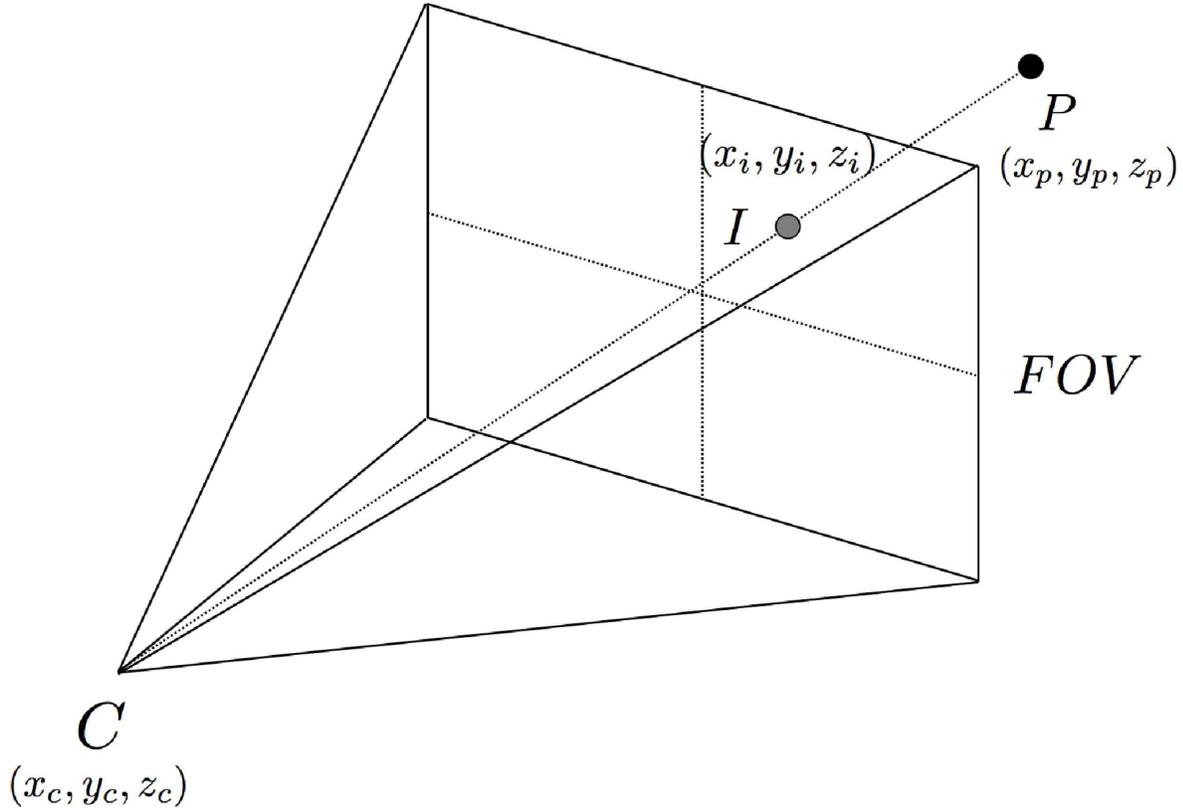}
\caption{The geometry of the position of the image $I$ of a particle $P$ as
seen from a camera $C$ looking towards a field of view FOV along the
$z$-axis for creating 3D perspective.}
\label{fig-perspective}
\end{figure}

The geometry of perspective for a single particle is given in Figure
\ref{fig-perspective}.  The pyramidal volume containing the system is
called the {\em frustrum}.  A camera $C$ at position $(x_c,y_c,z_c)$
is centered on the field of view (FOV) represented by a rectangle placed at
a distance $d=z_i - z_c$.
A ray traced from particle $P$ hits the FOV to create an image $I$
position $(x_i,y_i)$ with coordinates given by:
\begin{eqnarray}
x_i &=& x_c + \frac{(z_i - z_c)}{(z_p - z_c)}(x_p - x_c) \\
y_i &=& y_c + \frac{(z_i - z_c)}{(z_p - z_c)}(y_p - y_c)
\end{eqnarray}
One can create surface density maps by binning mass on the FOV
in the same way as parallel projections.  
The curvature of the FOV is ignored in this way and noticeable distortion
similar to wide angle photographs can be detected near the edges when the
angular width of the FOV is greater than approximately $90^\circ$.

In practice, a camera $C$ is specified by a position in space and an
orientation specified by a normal vector perpendicular to the image
plane pointed in the viewing direction.  For a distribution of particles, 
one first rotates the particles into the camera's frame of reference with
the viewing direction aligned with the $z$-axis and then take a ``picture"
by computing the surface density map long this line of sight.  
A simple way of obtaining the rotation matrix is through the use 
of quaternions \cite{sho85}.  
Quaternions are 4-valued quantities composed of a vector and scalar 
$(X,Y,Z,W)$ that compactly 
encode rotation through an angle $\theta$ about a given axis defined by 
the normal vector $(\hat{x},\hat{y},\hat{z})$.  The quaternion coordinates  are:

\begin{eqnarray}
X &=& \hat{x} \sin (\theta/2) \\
Y &=& \hat{y} \sin (\theta/2) \\
Z &=& \hat{z} \sin (\theta/2) \\
W &=& \cos(\theta/2)
\end{eqnarray}
The corresponding rotation matrix $M$ used to rotate particles about 
the defined axis is given by:

\begin{equation}
M = \left (
\begin{array}{rrr}
1 - 2Y^2 - 2Z^2 & 2XY - 2ZW & 2XZ + 2YW \\
2XY + 2ZW & 1 - 2X^2 - 2Z^2 & 2YZ - 2XW \\
2XZ - 2YW & 2YZ + 2XW & 1 - 2X^2 - 2Y^2
\end{array}
\right )
\end{equation}

One way to determine the rotation matrix for a camera pointing in the
direction of the normal vector $(\hat{x}_c,\hat{y}_c,\hat{z}_c)$ is to find
the axis in the $x-y$ plane that rotates through the angle
$\theta=\cos^{-1} \hat{z}_c$.  This axis lies along the line joining the
origin to $(\hat{y}_c,-\hat{x}_c)$ and it is straightforward to construct the
quaternion and resulting rotation matrix.

It is possible to move a camera through space with a variable position and
orientation to generate an animation through an image sequence of 
either a single N-body snapshot or series of evolving snapshots.  Rotations
and fly-throughs are common applications of moving cameras and we examine
some examples below.

\subsection{Full Sky and Dome Projections}

Full sky and dome projections are sometimes used to depict astronomical
datasets.
For example, an artistic panorama of the Milky Way in galactic coordinates
in an Aitoff projection based on real coordinates was created 
by Lundmark Observatory in the 1950's \cite{lun55}.  More recently, 
Axel Mellinger \citeyear{mel08} has created a similar 
panorama digitally from wide field
astrophotograpy.  Full sky maps of the cosmic microwave background
radiation have also been produced to present the primordial density
fluctuations recovered from the COBE and WMAP experiments \cite{smo92}.
Astronomy borrows transformations from cartography for mapping the objects
on the surface of a sphere to a flat display.
Two commonly used forms in astronomy are the Hammer-Aitoff (H-A) projection
for full sky presentations and the Azimuthal-Equidistant (AE) projections for
hemispheres.\footnote{see http://en.wikipedia.org/wiki/Map\_projection\#Equal-area}
The AE projection is used to create dome masters - flat
circular images - for digital planetarium content and resembles 
the images produced by a fish-eyed lens
in photography.

For the H-A projection, the general latitude and longitude coordinates
$(\lambda,\phi)$ are transformed into an $x-y$ pair that fills a 2:1
flat elliptical domain.  In astronomy, $(\lambda,\phi)$ are usually identified
with galactic coordinates $(l,b)$.  
This projections preserves area on transformation
so the relative angular size of objects is unchanged although there is some
distortion near the edges of the domain.  The transformation is:
\begin{eqnarray}
X &=& \frac{2\sqrt{2} \cos(\phi) \sin(\lambda/2)}{\sqrt{1 +
\cos(\phi)\cos(\lambda/2)}} \\
Y &=& \frac{\sqrt{2} \sin(\phi)}{\sqrt{1 + \cos(\phi)\cos(\lambda/2)}}
\end{eqnarray}
Particle positions  are transformed this way to a planar domain allowing
the creation of surface density maps.

To generate dome master images, 
it is easier to transform to spherical coordinates first.
If particles are centered on the view point and have spherical coordinates
$(r,\theta,\phi)$ then the azimuthal equidistant projection can be defined
simply as:

\begin{eqnarray}
X &=& \theta \cos \phi \\
Y &=& \theta \sin \phi
\end{eqnarray}

We can compute the spherical coordinates in the standard way from
$\theta = \cos^{-1} z$, $\cos \phi = x/\sqrt{x^2 + y^2}$ and $\sin \phi =
y/\sqrt{x^2 + y^2}$.  Normally, the radius is cut off at $\theta = \pi/2$ so
only one hemisphere is shown.  Digital planetariums transform images from this
projection typically resolved with $3600\times 3600$ pixels. 
Image processing software divides the dome masters 
into multiple images that are simultaneously projected onto a dome
to create a seamless image \cite{yu06}.

We will discuss the application of both of these projection methods 
in the case study of animating the collision of the Milky-Way and Andromeda
galaxies.

\subsection{Color and Brightness}

The surface density maps created in any projection are single valued and so
by default produce black and white images.  There is usually a large 
dynamic range in the values of projected densities so the default linear
mapping of density to a pixel with an 8-bit gray intensity value 
masks all detail.
This problem is overcome by ``stretching" the linear intensity through either a
power-law (gamma) or logarithmic transformation though a variety of
transformations are available depending on the dynamic range of the data.  
Similar transformations are applied to CCD images from modern telescopes to
create published images of galaxies and nebulae.

A variety of colour maps 
are available in standard image processing software like IDL and SAOImage
that map particular stretched
gray scale values to different colors.
While color maps are 
excellent ways of conveying physical detail and quantitative information
they sometimes are lacking in aesthetic appeal.  I present
some alternative ways of colorizing N-body simulations by assigning
individual colors to particle subsets.

In the case of simulated galaxies, a large fraction of the particles are
intended to represent stars or stellar populations so it makes sense 
to color them with the pallette of natural stellar hues.  
The color we perceive in an astronomical image comes from the
synthesis of a large number of sources with different intrinsic colors and
luminosities.  How do we calculate the final luminosity and color of a
image composed of a combination of different sources?
Our eyes perceive 
the spectral energy distributions (SEDs) of luminous source in terms of
color and brightness quantities that can be be reduced to three numbers.  
Modern computer monitors and digital images encode color with 24 bits (or
more) represented by the 3 8-bit values of red, green blue or RGB.  
RGB color is convenient to work but the net RGB values from a combination
of different sources cannot be determined additively.
A convenient color triplet for color synthesis is the set 
of CIE XYZ tristimulus values
\cite{poy96}. 
These values quantify the
human response to color through 3 weighting functions
$\overline{x}(\lambda)$, $\overline{y}(\lambda)$, and $\overline{z}(\lambda))$
that were determined from the statistics of experiments involving human
subjects.  The CIE XYZ values are calculated by integrating the product of
the SED $I(\lambda)$ with the weighting function over all wavelengths:

\begin{eqnarray}
X &=& \int_0^\infty I(\lambda) \overline{x}(\lambda) d\lambda \\
Y &=& \int_0^\infty I(\lambda) \overline{y}(\lambda) d\lambda \\
Z &=& \int_0^\infty I(\lambda) \overline{z}(\lambda) d\lambda
\end{eqnarray}

(Note that
the response drops to zero outside the visible range accessible to the
human eye).
The value Y is known as the {\em luminance} or net brightness in the
absence of color.  The normalized pair of values (x,y)  are known as the
{\em chromaticity coordinates} and are given by:
\begin{eqnarray}
x &=& \frac{X}{X+Y+Z} \\
y &=& \frac{Y}{X+Y+Z}
\end{eqnarray}
with inverse transformations

\begin{eqnarray}
X &=& \frac{x}{y}Y \\
Z &=& \frac{1-x-y}{y} Y
\end{eqnarray}
The {\em chromaticity coordinates} define a ``pure'' color in the 
absence of brightness.  An object's color can therefore be disentangled
from it's brightness.
Since images and displays used RGB color,
we need to convert XYZ triplets back into RGB.
There are various RGB systems some of which are device dependent. 
For example, one device independent definition system is Rec.~709 RGB
and conversion from XYZ to RGB can be done with the matrix transformation
under the assumption that $0<Y<1$.  

\begin{equation}
\left[ \begin{array}{r} R \\ G \\ B \end{array} \right ] =
\left[ 
\begin{array}{rrr}
3.240479 & -1.537150 & - 0.498535 \\
-0.969259 & 1.875992 & 0.041556 \\
0.055648 &-0.204043 & 1.057311
\end{array} 
\right ]
\left[ \begin{array}{r} X \\ Y \\ Z \end{array} \right ]
\end{equation}
Normally, RGB colors are constrained to lie between 0 and 1 so all RGB
values outside of this range are clipped to 0 or 1 depending on where
they exceed the range.  For our purposes, we do not necessarily need to
achieve a perfect true color representation of simulated galaxies
so the above transformation matrix is satisfactory though other ones can
also be used.

Since the XYZ color values are proportional 
to the emitted energy,  then
the net value of the XYZ color from combinations of luminous sources with
different SEDs is simply a sum of the independent XYZ values.  Therefore
when binning a system of particles (stars) with different brightnesses and
colors into pixels to create surface
brightness (density) maps, one can compute the net value of XYZ for the
pixel simply by summing over particles.  The chromaticity coordinates can
be determined to find the pure color of that pixel along with the luminance
Y.  Since the brightness profiles are usually sharply peaked in
astrophysical objects then luminance can be replaced by a
transformed value to bring out detail in the faint regions.  

A simple logarithmic transformation 
\begin{equation}
Y' = \log(1 + n Y/Y_{0})/\log(1 + n)
\end{equation}
enhances detail well
where $n$ is a free parameter with $n=10$ being a good choice and $Y_{0}$
is a reference intensity near the maximum value in the image.  

Another tranformation is a power-law or gamma transformation is:

\begin{equation}
Y' = (Y/Y_0)^\gamma
\end{equation}
where $\gamma$ is a power-law typically less than 1 and $Y_0$ again is a
threshold luminance usually near the maximum value in a map.  
Pixels can then be colored according to the
chromaticity coordinate determined during the creation of the map.
Finally, after these transformation images can be
improved further by adjusting the values of the brightness and contrast
using a variety of tools.  The degree of adjustment will depend on the
dynamic range of the luminance in the image.

To summarize, we can colorize N-body surface density maps using 
a simple color map that maps a gray value to a specific color
or through color synthesis using CIE XYZ tristimulus values.
With color synthesis, we assign individual stars their own unique
colors and luminance expressed as an XYZ triplet.  we then generate
three independent surface density maps for each tristimulus value 
X, Y, and Z or alternatively we generate surface density maps for different
subsets of particles assigned different colors.  Since the XYZ color values
are proportional to emitted energy then we can sum over particles when
binning in pixels to create an image.  We determine the pure color 
of each pixel from the chromaticity coordinates (x,y) derived from each
triplet.  We can also stretch the luminance Y in each pixel stretched with an
appropriate transformation to reveal faint detail and then color with
the derived chromaticity.  The RGB colors of each pixel 
calculated with the linear transformation above create the final color
image.  We discuss applications of color synthesis below in case studies of
star clusters and galaxy collision simulations.

%
%
%
%

\subsection{MYRIAD: An N-body visualization software library}

Many of the ideas discussed above are the basic elements of standard 3D 
rendering packages mentioned in \S 2. 
Rendering astrophysical particles in the absence of an
absorbing medium is basically straightforward since the particles are
simply luminous, transparent points obviating the need for 
complex raytracing. 
The main difficulty at the current time is dealing with datasets
that now contain billions of particles.  While rendering single images is
not difficult, creating animations of evolving datasets with fine time
resolution can be cumbersome. Large file sizes make saving numerous
N-body snapshots prohibitive since storage space can be limited and
I/O operations can be slow. 

One solution to this problem is to integrate a visualization software library
with the N-body code itself.  The task of rendering many types of
images is small in
comparison to computing the gravitational (and hydrodynamic) forces
in simulations
so images from multiple cameras at different positions and orientations
can be taken at small time intervals.  Large N-body simulations today are
done on parallel supercomputers and most algorithms distribute particles
among the computing nodes almost equally.  Raw surface density maps 
from a static or moving  camera position can be created in parallel 
on different processors.  The
final image can be synthesized by summing over partial images generated
independently on computing nodes.  

I have developed a software
package called  MYRIAD that creates raw binned
surface density maps from particle distributions as orthographic
projections, perspective images from static or moving cameras, 
full sky or full dome  projections and stereoscopic images.
Post-processing software is used to apply color synthesis methods by 
blending raw images for different ranges of particles assumed to have
different colors or different color bands
as well as applying convolutions with different point
spread functions to either simply smooth images or add star-like
characteristics for bright pixels.
The rendering software has been integrated with the
parallel N-body codes PARTREE \cite{dub96} and GOTPM
\cite{dub04} so that image sequences are created during simulation runs.
The package can also be used to generate single images or
sequences on a single data snapshot through a command line programme.
In this way, it has been possible to create sequences
with thousands of images in large simulations without the need to save
the same number of N-body snapshots.

I will now discuss specific applications of visualization in both
galactic dynamics and cosmology that have used MYRIAD to create images and
animations.

%
%
%
%
%

\section{Case Studies}

The dynamics of gravitating systems of particles
representing galaxies and the cosmological large-scale structure 
are not only scientifically interesting to astronomers but also a 
source of beauty and wonder to a more general audience.  
Over the past few years, I have been involved in
a project called GRAVITAS that attempts to express both the
scientific complexity and artistic beauty of Newtonian gravity and dynamics
revealed through N-body simulation.
Animations of star clusters, spiral galaxies, galaxy interactions, galaxy
clusters and the large-scale structure are rendered in a variety of ways 
with fine time-resolution.   The animations are accompanied by original music
by modern classical composer John Kameel Farah who
crafted pieces to reflect both the evolving imagery and scientific
ideas in the animations.
All of the animations are available in a variety of formats including DVD
and free downloads at resolutions up to high-definition can be obtained
at our website
\cite{gra08}.

\subsection{Cosmological Simulations}

Cosmological simulations reveal the evolving gravitational framework of the
universe as seen in dark matter.  They begin when the universe is nearly
homogeneous and featureless and show the growth of complex structure
through the hierachical collapse of primordial density fluctuations.  Small
objects condense first and merge to form larger objects that become the
dark matter halos of galaxies and galaxy clusters.

\begin{figure}
\centering
\includegraphics[width=15.5cm]{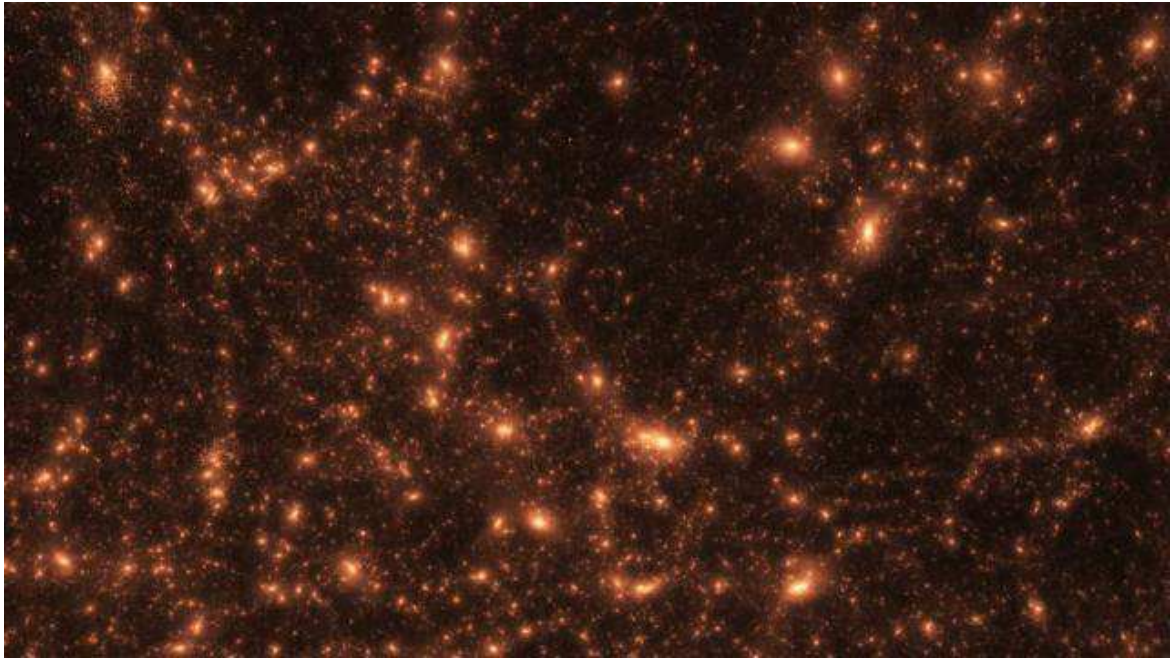}
\caption{The final snapshot from the animation ``Dark Universe" - a
fly-through of an evolving cosmological dark matter simulation.  At this
stage, the universe is filled with dark halos - the structures underlying
galaxies and galaxy clusters.}
\label{fig-darkuniverse}
\end{figure}

The first application involved the rendering of a flight through a
cosmological dark matter simulation (Fig. \ref{fig-darkuniverse}).  
I used the code GOTPM code to simulate the formation
of structure within a periodic cube 100 $h^{-1}$ Mpc on a side with initial
conditions determined by the currently favoured parameters of the $\Lambda$-CDM
model using $1024^3$ particles.    The simulation was computed on a
parallel cluster using 256 processors 
at McMaster University funded by the SHARCNET collaboration.
In cosmological simulations, 
the particles are confined to a periodic cube and the equations of motion
are integrated in co-moving coordinates to null out the cosmic expansion.
To emphasize the growth of structure, N-body images are rendered in
co-moving coordinates at each of the 5000 timesteps.
 
The image parameters, camera coordinates and viewing direction 
are specified through a data file provided to the MYRIAD software.  
The camera was set on a straight line course at constant speed in co-moving
coordinates so that it would fly through the cube twice over the course of
the run.  We replicated one layer of the cube so
that the camera would never get closer than one cube edge length from
the edge of the particle distribution.  In the end, 
the renderer looped through 27 billion particles at each step 
to check which particles were in
the field of view for creation of the surface density maps.  The camera's
field of view was set to one radian along the widest dimension in frames
that were rendered at Cinematic 4K resolution 
(4096x2160 pixels).  A surface density
map was created from the moving camera at each integration timestep for a
total of 5000 images.  The gray images were smoothed slightly with a
Gaussian PSF and then enhanced  and
colorized with a hue that changed gradually from blue to deep
red over the course of the simulation as an artistic 
way to convey the aging of the universe.

A single snapshot from the last frame representing the
system at the current redshift $z=0$ is shown in Fig.~\ref{fig-darkuniverse}
and reveals the dark halos from the scale of galaxies to galaxy clusters.
The accompanying animation shows the evolution 
of dark halos from a homogeneous beginning through 
the hierachical collapse and merger of smaller
structures emerging from the primordial density fluctuations.
At this resolution, we can see substructure within the largest dark halos 
in the form of orbiting dark matter satellites.

\subsection{Nightfall}

Globular star clusters are one of the purest expressions of 
Newtonian gravitational dynamics in the universe.
They are also the
oldest objects in the galaxy and have a distinctive distribution of stars
reflecting an old population that is visually dominated by a 
red giants and blue horizontal branch stars as can been seen in images
of globulars
from the Hubble space telescope.
These objects have been fascinating to astronomers and amateurs alike.
Asimov \citeyear{asi41}
in his classic pulp science fiction story ``Nightfall" tells the
tale of a civilization on a planet in a multiple star system in a globular
cluster.
The people of this world are driven to insanity when all of their suns 
set for the first time in thousands of years
revealing a night sky filled with tens of thousands of bright stars. 
Robert Burnham, Jr. \citeyear{bur78} 
speculates on how the sky might appear inside the
Hercules globular cluster M13:

\begin{quotation}
"The appearance of the heavens from a point within the Hercules Cluster
would be a spectacle of incomparable splendor; the heavens would be filled
with uncountable numbers of blazing stars which would dwarf our own Sirius
and Canopus to insignificance.  Many thousands of stars ranging in
brilliance between Venus and the full moon would be continually visible, so
that there would be no real night at all on a planet in a globular
cluster.
\end{quotation}

Inspired by real pictures and these speculations, I created a
realistic and dynamic view of a globular cluster over a few hundred orbital
times using the treecode PARTREE.   
For the particle distribution, I use a 1 million particle 
King model \cite{kin66} - one of the standard, isotropic distribution
functions that are used to model globular clusters.
I assume that each particle is just a star of a given visual magnitude $V$
and photometric color $B-V$ using a table of data for M15.  
The color index $B-V$ can be mapped to an effective temperature 
to estimate a stellar color.  
The tristimulus color indices of the black-body radiator at the
effective temperature for each star were computed by integrating over 
the blackbody intensity function and then normalized to 
stellar luminosity measured by the absolute magnitude $V$.  
This is an approximation
but gives colors that are close to the true values.  

\begin{figure}
\centering
\includegraphics[width=15.5cm]{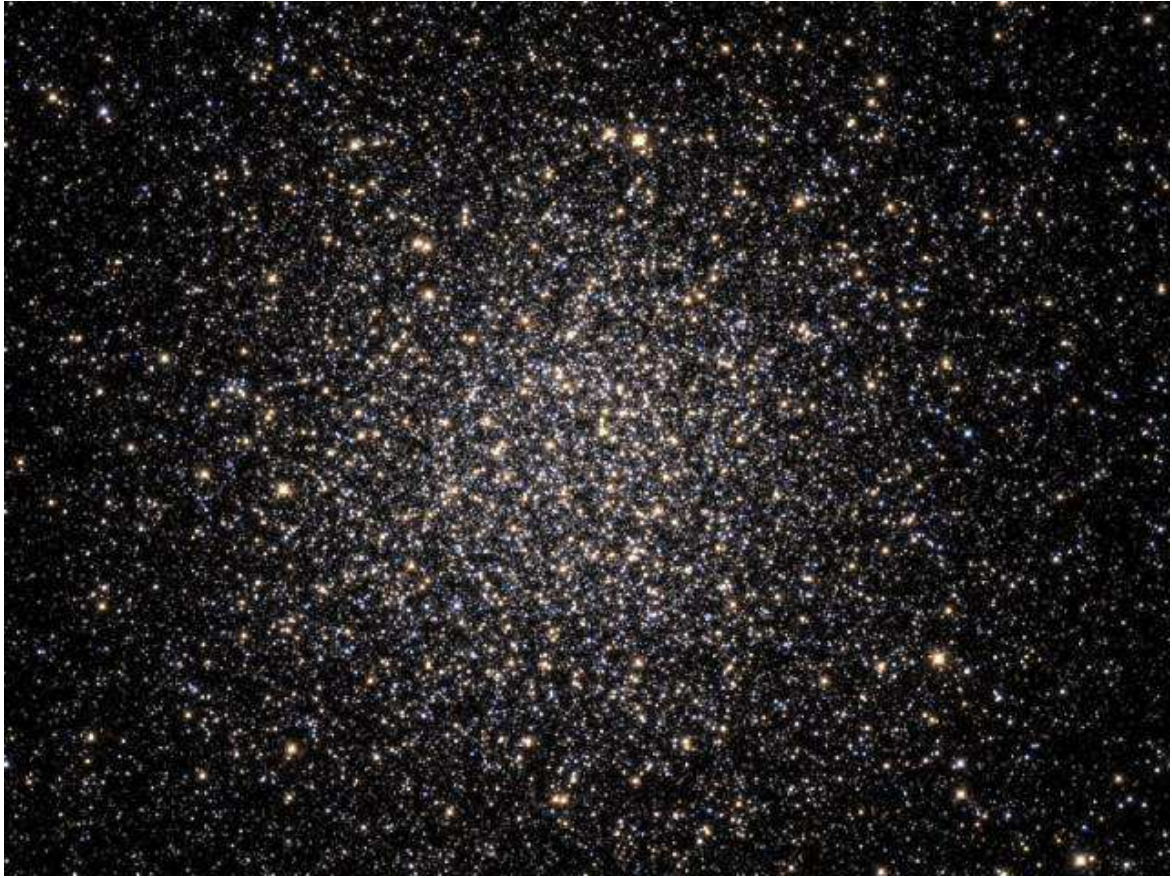}
\caption{A snapshot from the animation ``Nightfall"  a photorealistic
depiction of an inspiralling flight into a globular cluster.}
\label{fig-nightfall}
\end{figure}

At each step of the
simulation a surface density map for each of the three tristimulus
values was created for the particle distribution.  In these maps,
individual lit pixels tended to represent single stars.
Each map is also
convolved with a PSF generated by the Tinytim package \cite{tinytim} used
to characterize the properties of the PSF in the Hubble Space Telescope
(HST) wide field cameras.  A typical HST camera PSF has 
diffraction spikes that become
apparent for bright stars and so added another dimension of realism to the
images (Fig.~\ref{fig-nightfall}).  
The apparent brightness of stars was also allowed to vary depending on the
distance from the camera following a $1/r^2$ intensity law so that
approaching or receding stars would brighten or dim.  This effect added
a further sense of depth to the animation.
Finally, for a camera path we choose an inspiralling trajectory
that keeps the camera directed to the center of the cluster starting from a
distance that reveals the crowded cluster core.
In this way, we get a 3D sense of the structure of the
cluster and also witness the dynamics of the stars as we approach the
center of the stellar distribution.

\subsection{Milky Way - Andromeda Collision}

Galaxy collisions are dramatic and visually stunning examples of
gravitational dynamics on the large scale.  The mutual tidal forces of
closely interacting disk systems excite 2-armed spiral structure that throw
off long tails of stars and gas.  The loss of orbital energy to random
internal motions lead to the merger of the disks and transformation into an
elliptical galaxy.  The unravelling of the disk systems leads also
patterns of shells and ripples that are themselves another reflection of
the physical beauty of phase-mixing in collisionless particle systems.
The desire to simulate  the complex dynamics of
galaxy interactions has been
one the main motivations for developing new N-body methods and
visualization methods and the animations I describe below continue in this
tradition.

One of the more exciting popular astronomical ``facts'' to emerge in the 
last decade is that our own galaxy, the Milky-Way (MW) seems to be on a
collision course with our neighbour spiral galaxy Andromeda (M31)
\cite{dub06}.  This idea originates from the original interpretation of the
MW-M31 system as a bound pair of orbiting galaxies 
by Kahn \& Woltjer \citeyear{kah59}
in the so-called timing argument.  More recently, modelling of motions in
the Local Group of galaxies suggest that this collision may occur within
the next few billion years \cite{ray89}.  This is a compelling timescale
since most eschatological discussions of the Earth's fate are focussed on the
influence of the evolving and warming Sun and it's eventual transformation 
into a red giant - a process that occurs on a somewhat longer timescale of
5 billion years.   Before that final end, the MW may very well
collide and merge with M31 creating a new list of disaster
scenarios to explore for the Earth's possible demise.  A simulation of
the MW-M31 collision is certainly worth a look both as a specific example
of a galaxy collision in action as well as compelling story that may
involve the fate of the Earth and solar system and any life that might
still be hanging around in a few billion years.  Cox \& Loeb
\citeyear{cox08} have even written
a scientific paper recently that claims predictions for the eventual
fate of the Sun based on model-dependent assumptions.
I discuss here my own simulation which is more speculative
but does incorporate known constraints on the positions, motions and
mass distributions of the two galaxies.  The idea is not to quantify the
statistics of our eventual fate but rather to explore one detailed 
scenario to give a sense of what may happen and what it might look like
to an Earth observer.

The initial conditions for the collision are described in Dubinski \etal
\cite{dub96} in a discussion of the formation of tidal tails in galaxy
interactions.  The galaxy
orientations, relative distance and positions are defined here for specific
galactic models that have dimensions, rotation curves, and mass
distributions similar to the Milky-Way and M31.  These original simulations
only used around 30K particles but the models have been re-run with
more particles over the years in various demonstration calculations
\cite{nor96,dub06} with the recent largest version containing more 
than 300M particles on a parallel supercomputer in Toronto in 2003
\cite{dub03}.

\subsubsection{Spiral Metamorphosis}

\begin{figure}
\centering
\includegraphics[width=15.5cm]{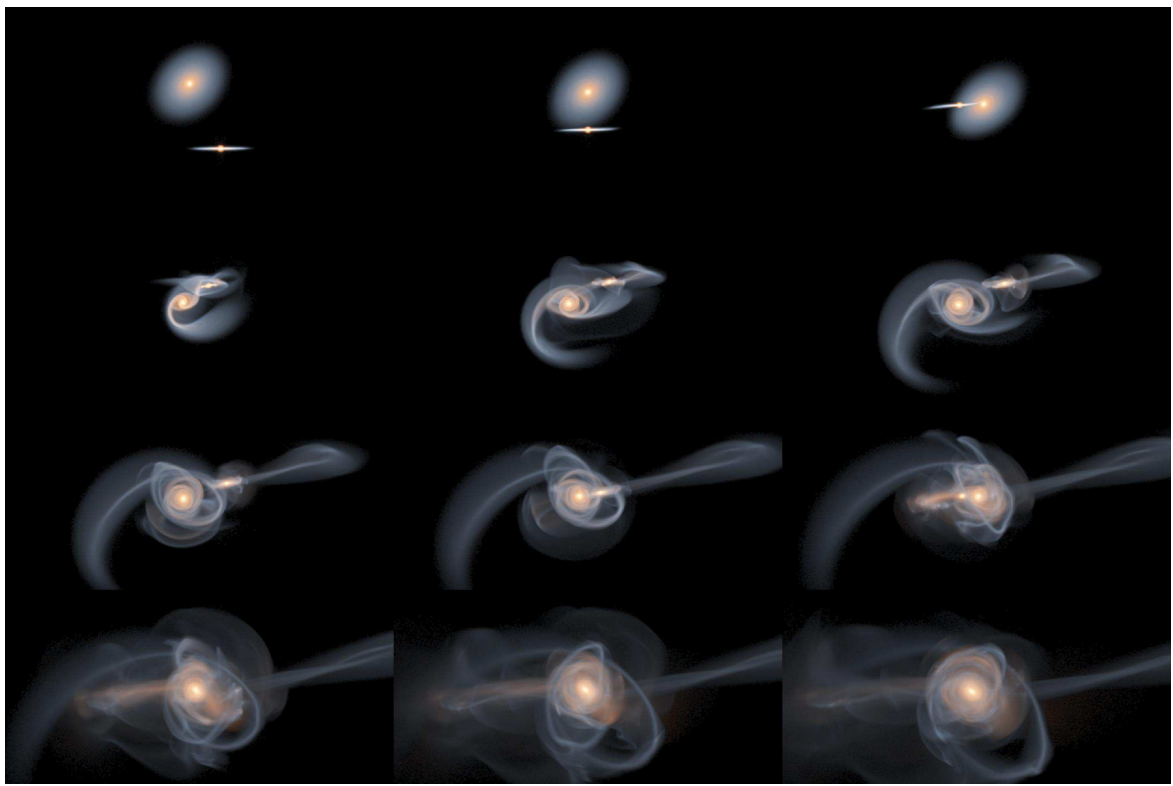}
\caption{An image sequence from the animation ``Spiral Metamorphosis" that
depicts the merger of the Milky-Way and Andromeda galaxies.  The time
between images is about 170 million years and shows the view looking down
on Andromeda.  The tidal interaction excites spiral structure
and throws off long tidal tails and bridges from the two galaxies.  The
galaxies eventually merge revealing a complex pattern of shells and
streams as they transform into an elliptical galaxy.}
\label{fig-mw}
\end{figure}

In this animation, the collision is depicted from a static camera placed
at a distance of 320 kpc from orthogonal directions aligned with the
galactic plane.  Two versions of the animation were created. 
The first animation uses
300M particles (Fig. \ref{fig-mw}) 
and presents two orthogonal projections with a
face-on and edge-on view of the Milky Way.  A second
smaller 30M particle simulation was used 
to generate a stereoscopic rendering that was converted
to a red-cyan anaglyph for viewing with colored stereo
glasses.  For stereoscopic rendering, one simply takes pictures of the same
view from two cameras separated by a small angle (see Bourke \citeyear{bou07}
or McAllister \citeyear{mca93} 
for an explanation of techniques of stereoscopy).
A unique feature of these animations is the fine time resolution
compared with other renditions in recent years.  
Each viewpoint is rendered with 5300 steps
covering 2.3 billion years or 440K years per step starting a few hundred
million years before the first close passage of the two galaxies.  
At 30 frames per second,
time passes at the rate of 13 million years per second.  At this rate, the
animation of the collision takes nearly 4 minutes
revealing great detail while enhancing the majesty of the event.

Only the particles representing stars were rendered in these animations
with the dark matter left invisible.  I
used a simple two color scheme to represent old (red) and blue (young)
stellar populations and the raw images were convolved with a Gaussian PSF.  
The chosen tristimulus colors were generated 
from the SED of young and old populations generate using Bruzual \& Charlot
\citeyear{bru93} models obtained from Bob Abraham \citeyear{abr02}.
The derived colors are close to those
depicted in ``true" color Hubble images 
of interacting galaxies like NGC 4676 ``The Mice".  Bulge stars were
colored red while the disk stars were colored both red and blue with the
probability of being red increasing towards the center of the Galaxy. I
assigned an exponential probability as a function of gravitational binding
energy within the galaxy potential with a functional scale length that
would show a gradual color transition from red to blue from the inner to
outer disk.
In this way, the initial galaxy models have appearances
similar to true color images of spirals like M31.  
As the galaxies interact
and eventually merge, the blending of the two stellar colors create 
an interesting result.

At this high resolution, one should note that the initial model galaxies
are extremely smooth and gravitational instabilities that generate 
spiral structure from the Poisson noise are greatly suppressed. 
These animations present an idealized depiction of a smooth stellar
distribution and are missing the knotty structure of real spiral galaxies 
containing clumps of gas and new star formation regions.  The obscuring
dust is also missing so the familiar dust lanes seen in real galaxies is
not apparent.  The current space show at the Hayden Planetarium in New York
presents a depiction of the MW-M31 collision that includes gas and 
star formation along with dust obscuration using a few million particles.

\subsubsection{Future Sky}

\begin{figure}
\centering
\includegraphics[height=8.0in]{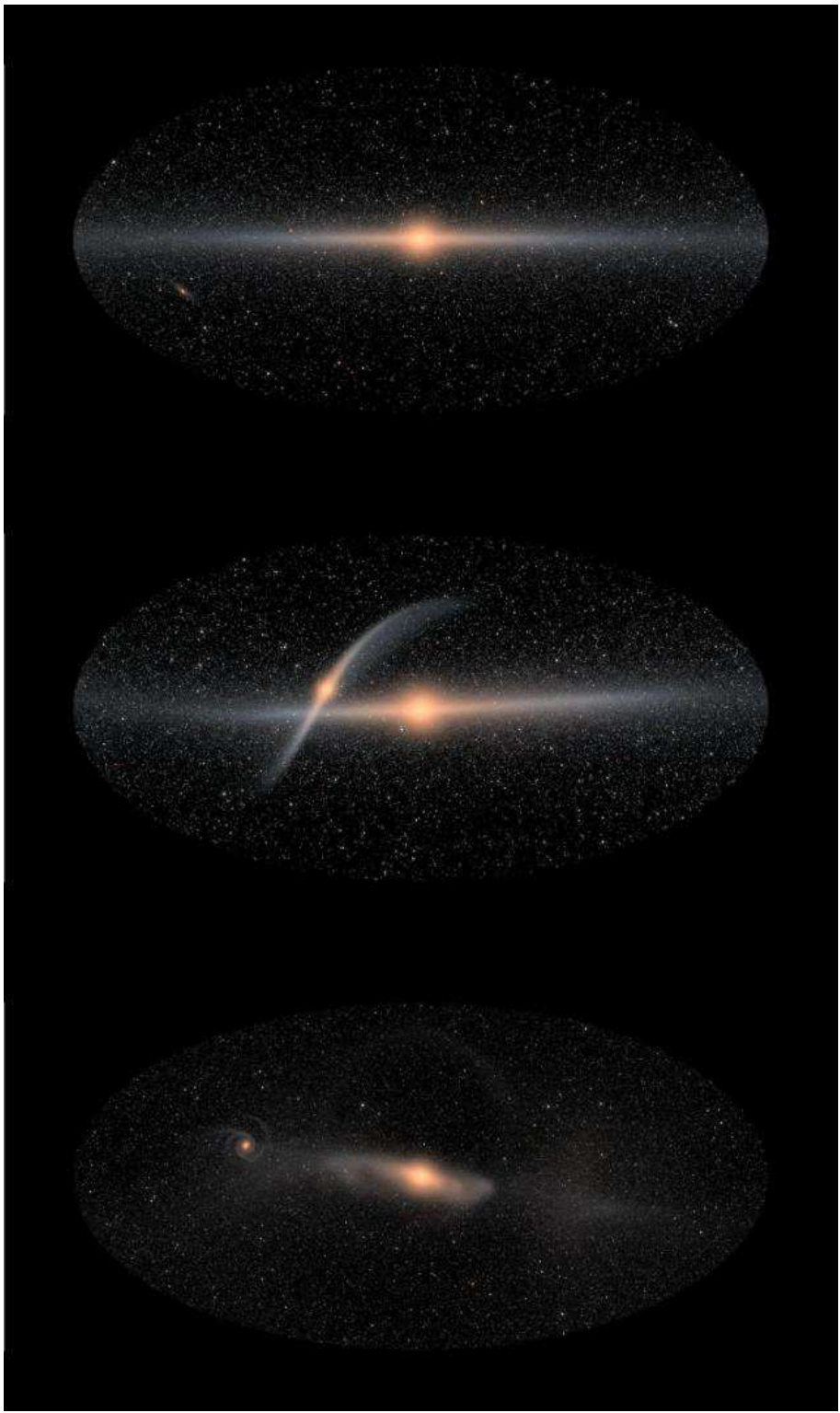}
\caption{Three frames from the animation ``Future Sky" showing the
initial transformation of the sky during the Milky-Way Andromeda merger
from the perspective of the Sun.  The 3 frames depict 1) the initial
unperturbed Milky-Way with Andromeda appearing as a small ellipse in the
left beneath the galactic plane 2) the view when Andromeda crosses the
plane of the Milky Way during closes approach 3) the view shortly after
when Andromeda recedes leaving a disturbed Milky-Way.}
\label{fig-futuresky}
\end{figure}

\begin{figure}
\centering
\includegraphics[width=15.5cm]{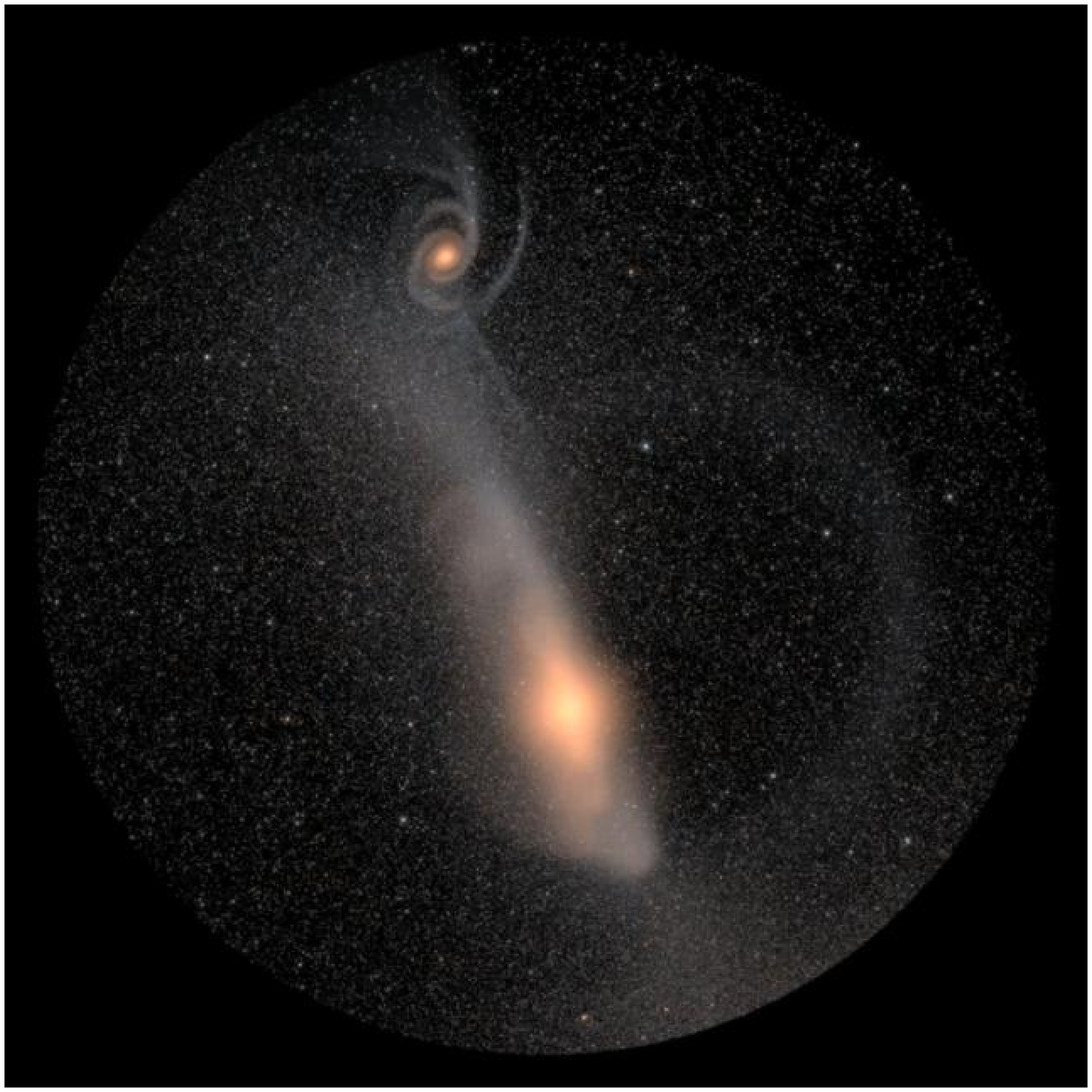}
\caption{A dome master image from the animation ``Future Sky" showing the
band of the Milky-Way, the galactic bulge and the perturbed Andromeda
galaxy with excited spiral structure.  Images like these are used to
generate projections onto the domes of digital planetariums.}
\label{fig-dome}
\end{figure}

This animation uses the full sky Hammer-Aitoff projection 
and the Azimuthal-Equidistant projection for planetatrium dome masters
to present the
MW-M31 collision from the perspective of the Sun.  This animation was
presented at the Computer Animation Festival at SIGGRAPH 2006 \cite{dub06a}
and the dome
version was also used for a special presentation at the Gates Planetarium
in 2006 (see Figures \ref{fig-futuresky} and \ref{fig-dome}).
One particle on a circular
orbit at the solar distance of 8 kpc from the center of the MW model is labelled
as the Sun and acts as the platform for the moving camera.  
As the system evolves, the position of this particle is provided 
to the image rendering
software as the origin for the full sky projections.  
Initially, the line from the
Sun to center of the bulge represents the X-axis in galactic coordinates.
We assume future observers will continue to use this convention and define
a galactic coordinate system based on this axis and the normal vector to
the initial Galactic plane.
To track the bulge, we add another
particle to the simulation with the mass of a supermassive blackhole in the
center of the bulge of the MW.  This particle is also given a large
gravitational softening length to prevent undesirable dynamical evolution
of the bulge.  Because of the effect of dynamical friction
this particle remains within the dense center of the bulge and therefore
acts as a bulge center reference point.
We re-orient the particles at each step into a new galactic coordinate
system that is defined by the X-axis of the line joining the sun to 
the blackhole at the bulge centre and the
original normal vector to the MW galactic plane.  In this way, the bulge
tracked by the blackhole always lies along the central meridian in the H-A
projection and initially the band of light representing the midplane of the
Milky-way remains aligned with the equator.  At later times, when the Sun
is thrown out of the plane, the bulge begins to show apparent oscillatory
motion along the meridian.  The bulge also grows and diminishes in size 
as the Sun rides on a plunging radial orbit that takes it through the
center of the bulge.

As a small trick, we use the PSF from the Hubble Space Telescope even
though the the images show the full sky.
In this way, very bright nearby stars have 
both a noticeably larger diameter and spikes enhancing the impression of
brightness and proximity while dim stars are more point like.
One final adjustment is the transformation of the linear intensity using
logarithmic transformation.  In this way, faint interesting
details are enhanced and can be seen beside the brightest nearby stars.

\section{Conclusions}

The visualization of N-body systems has evolved from simple dotted figures
a few decades ago to detailed computer animations 
with billions of particles incorporating the latest developments 
in 3D computer graphics methods.  Numerical astrophysics is an
experimental science and the images and animations that are created from
simulations are valuable tools for interpreting the results and comparing
to observations of the real universe.  A well-crafted visualization is 
a powerful way of conveying detailed physical processes to a general
audience and greatly enriches understanding of phenomena.

I have presented some specialized methods for rendering images and 
animations of large N-body simulations that build on basic principles 
of 3D computer graphics.  The techniques described are straightforward 
and relatively easy to implement and many groups have replicated the
effort in different software.  The main new development here is
is the ability to render animations with fine time resolution 
efficiently in parallel supercomputer simulations by integrating 
the visualization library MYRIAD with an N-body code.  I have provided case
studies of creation of a selection of animations from the GRAVITAS project
that explore different features of the library.

As the resolution of simulations increase, model galaxies are
beginning to look more and more like the real thing.  
The iconic Hubble color images of galaxies are
good working definitions of an expression of the visual reality 
of the universe.
A challenging goal for the near term might then be to create
photo-realistic animations of the dynamics of galaxies that resemble
the Hubble images using our best N-body models of galaxies containing gas,
stars and dark matter.  
The methods I describe here only describe renderings of
purely stellar galaxies while the dust lanes and star forming regions of
real spiral galaxies are ignored.  
Simulations of galaxies including gas dynamics, star formation and AGN
feedback are being
carried out by different groups now (eg., 
Di Matteo \etal \citeyear{dim05},
Governato \etal \citeyear{gov08})
and resolution is increasing as well though
the extra computational cost of including gas have kept simulation particle
numbers to a few million.
The animation of the MW-M31 galaxy collision the planetarium show ``Cosmic
Collisions'' created by the Hayden Planetarium 
is a beginning in that direction of simultaneously
rendering stars and the obscuring affect of interstellar dust in an animation.
Another factor of 10 to 100 in particle numbers is probably required 
to approach a truly realistic view.

\section*{Acknowledgments}

I acknowledge the Canadian Institute for Theoretical Astrophysics (CITA)
and the Shared Hierarchical Academic Research Network (SHARCNET) for
providing supercomputer time for simulations and visualizations.   Research
funding was provided by NSERC.  I also acknowledge helpful discussions 
with Bob Abraham, Ka Chun Yu, Frank Summers, Alar Toomre, and Rachid Sunyaev.



\section*{References}

\bibliographystyle{jphysicsB}
\bibliography{refs}

\end{document}